# Resolution Limit in Microsphere-Assisted Super-Resolution Microscopy

ARLEN R. BEKIROV[1,*], ZENGBO WANG[2], NINA A. LYSTSEVA[1], BORIS S. LUK'YANCHUK[1], AND ANDREY A. FEDYANIN[1]

[1] *Faculty of Physics, Lomonosov Moscow State University, Moscow 119991, Russia*
[2] *School of Computer Science and Engineering, Bangor University, Bangor, Gwynedd, LL57 1UT, UK*
*bekirovar@my.msu.ru*

**Abstract:** Visualization in the virtual image formed by dielectric microparticles has been shown to enable the distinction of objects that remain indistinguishable under direct observation. We perform the resolution analysis based on a full two-dimensional simulation of optical image formation taking into account the diffraction of partially coherent light on the microparticle and the objects under study. The oscillating nature of optical resolution is demonstrated depending on the size of the microparticle. The presence of strong resonances is observed in both transmission and reflection modes. It is shown that as the size of the object decreases, the optical resolution tends to the classical limit. An analytical estimate for the resolution criterion in microsphere-assisted imaging is presented.

## 1. Introduction

Examining objects with the help of dielectric microparticles allows one to resolve structures beyond the diffraction limit [1]. Various theoretical approaches have been proposed to explain this phenomenon [2-14]. Among them, some works are focused on modeling the propagation of radiation from a light source through the object and the microsphere up to the subsequent image formation [9, 10, 14]. In these models, the objects are not point sources but have finite dimensions. Due to computational complexity, such calculations are typically limited to the two-dimensional case. Experimental data confirm the super-resolution effect in this configuration [15]. However, the crucial question of whether a fundamental resolution limit exists remains unresolved. We examine the minimum achievable resolution in virtual imaging formed by microparticles and its dependence on particle size in both reflection and transmission.

Recently, we proposed a simulation method for microsphere-assisted super-resolution phenomena based on the FDTD (Finite-Difference Time-Domain) approach, which demonstrated its feasibility [14]. However, this method is computationally demanding. For a monochromatic light source with wavelength $\lambda$, the FEM (Finite Element Method) is more efficient and accurate. Therefore, the calculations are performed using the FEM method implemented in the Matlab Partial Differential Equation Toolbox, utilizing the specialized "electromagnetic" class of the "harmonic" type.

## 2. Results

The system under study consists of a substrate and a dielectric microcylinder with a refractive index of $n = 1.46$. The sample is placed between the substrate and the microparticle. In the reflection mode, it consists of two rectangular metallic objects with a width of $0.25\,\lambda$; in the transmission mode, the structure is represented by slits of the same dimensions in a metallic screen. The system is illuminated according to the Köhler scheme. Further details are provided in the *Methods* section.

To determine the optical resolution, we performed a series of calculations for a fixed geometry system while varying the distance between objects. The resolution was determined using the classical bisection method.

1. Define the lower and upper resolution bounds as distances $S_1$ and $S_2$ where the objects are unresolved and resolved, respectively. By default, the resolution is assumed to be $S_2$.
2. Next, we check whether the objects are distinguishable at the midpoint distance $S_{12}=(S_1+S_2)/2$. If the objects are resolvable, we update the upper bound to $S_2=S_{12}$; otherwise, we set the lower bound to $S_1=S_{12}$.
3. Repeat step 2 until the difference between the upper and lower bounds becomes smaller than the predefined accuracy, $(S_2-S_1)<$accuracy. We set the accuracy to $0.01\lambda$.

As the initial values, we set $S_1=0$ and $S_2=0.3\lambda$. If the objects remain unresolved at $S_2=0.3\lambda$, the upper bound is increased by $0.1\lambda$ until the objects become distinguishable.

In some cases, the resolution condition is satisfied even with zero separation ($S_2=0$), indicating that the sphere does not form an image that defines the object's geometry. In such cases, the resolution is considered indeterminate. These cases are discussed in more detail later.

Fig. 1 illustrates the dependence of optical resolution on the size of the microparticle in transmission mode and reflection mode. To visualize the resonance peaks, we added more calculation points for the reflection mode, while in all other cases, the calculation grid remained standard $R=(4:0.005:5)\lambda$. Fig. 1 also shows the resolution in free space, which slightly differs from $\lambda/2$ and is approximately $0.55\lambda$. This is explained by two factors: the objects are not point sources and the coherence radius is not zero.

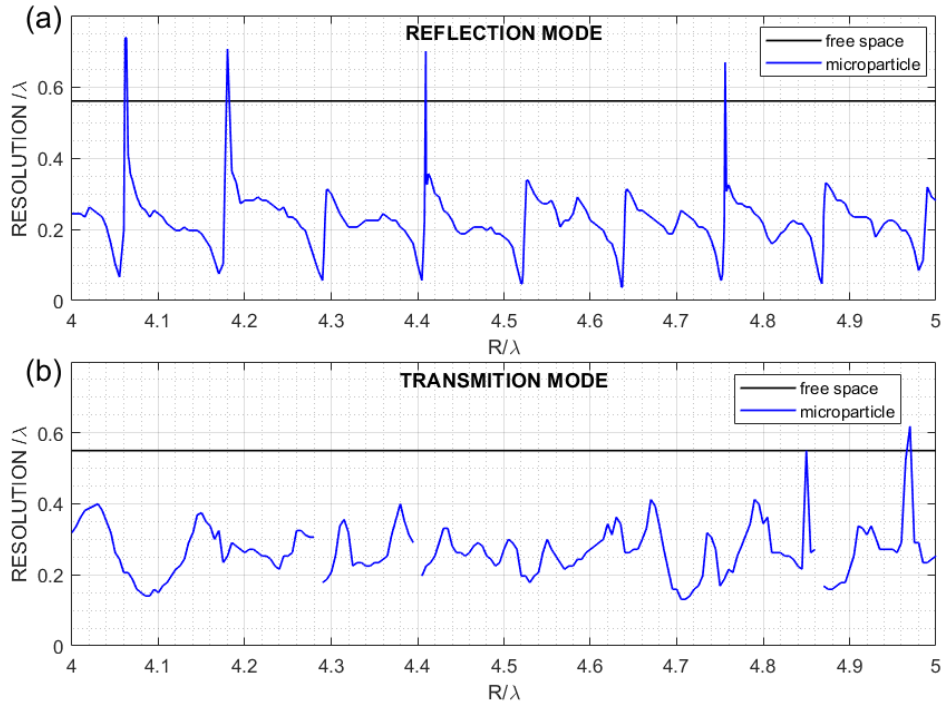

**Fig. 1.** Dependence of the optical resolution in the virtual image on the size of the microparticle: reflection mode (a), transmission mode (b). The black solid lines indicate the resolution in free space without the microparticle. At the discontinuity points in the graph for the transmission mode, the resolution is undefined because the slits are distinguishable (i.e., the resolution condition is satisfied) at zero distance. The resolution in free space matches within the error margin ($0.01\lambda$): $0.57\lambda$ for the transmission mode and $0.56\lambda$ for the reflection mode. To visualize the resonance peaks, more calculation points were added for the reflection mode (a).

The calculations showed that the resolution in free space, i.e., in the real image, is approximately the same for both reflection and transmission geometries. However, for the virtual image, i.e., when a microparticle is present, it differs significantly. For the most cases, the optical resolution surpasses the free-space limit. We also considered critical illumination. In this case, no significant changes were observed in the reflection mode, whereas in the transmission mode, the results differed considerably. Under critical illumination, the graph exhibits rapid changes, which are attributed to coherent effects.

The graphs exhibit two types of distinctive features: the areas of sharp resolution changes in Fig. 1(a), while Fig. 1(b) shows discontinuities. The physical origin of these features lies in the resonant field enhancement inside the particle, where the amplitude significantly increases, leading to the excitation of a whispering gallery mode. In this regime, the field inside the particle forms distinct bright maxima near its boundary (see Fig. 2(a)). The characteristic width of these sharp resolution changes or discontinuities is approximately $\Delta(R/\lambda) = 0.05\lambda$.

Fig. 2 shows the near-field distribution inside the particle and the corresponding image field near the sharp resolution changing point at $R = 4.06\lambda$. The case of minimal resolution is illustrated in Fig. 2(a), where the formal resolution reaches an extremely small value of $0.06\lambda$ at $R = 4.055\lambda$. However, when compared to the case at $R = 4.06\lambda$, where the resolution is $d = 0.2\lambda$, the image fields appear nearly identical.

Thus, despite the particle separation is differing by a factor of three, the distance between maxima in the image field remains nearly the same. This suggests that the image field does not precisely convey the geometry and spacing of the objects but rather indicates whether one or two particles are located beneath the microsphere. Such behavior is characteristic of the excitation of antisymmetric modes, a phenomenon noted in Ref. [16].

The cause of the sharp resolution reduction in Fig. 2(c)–(d) can also be understood: it arises due to the formation of a pronounced maximum above the geometric focus. As a result, the formal resolution condition is no longer met, even though two distinct stripes can still be observed below this maximum.

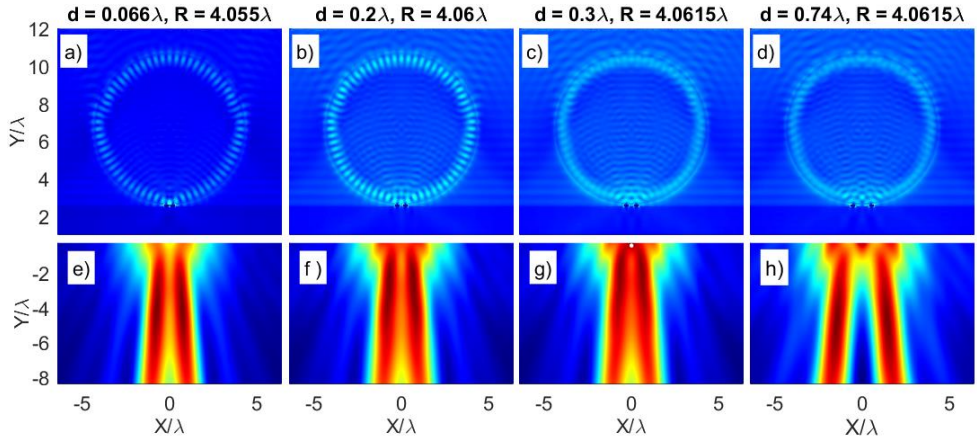

**Fig. 2.** Near-field distributions in reflection mode around sharp resolution changing at $R/\lambda = 4.06$. The white dot in (g) indicates the maximum field. Due to the emergence of this maximum, the optical resolution significantly deteriorates to $0.74\lambda$. The maximum values of the near fields for (a)–(d) are related as 1:0.6:0.6:0.6. The bright stripes at the microparticle boundary in (a)-(b) indicate the excitation of a whispering gallery mode.

Consider the break points in Fig. 1(b). The resolution at these points is undefined because the resolution condition is satisfied even at zero distance between the slits. We found such cases only in the transmission mode. Let us examine this case in more detail using the break point at $R/\lambda = 4.4$ as an example. At the first glance, it might seem that the resolution condition would continue to hold as the slit distance increases, but this is not the case. When calculated for a distance of $d = 0.1\lambda$, these conditions are no longer met, and this trend persists up to $d = 0.2\lambda$, a value that can be estimated by interpolation from the graph. Therefore, with some reservations, this value can be considered the optical resolution in this case. Fig. 3 illustrates this behavior in more detail. This example highlights the need to develop more general criteria for defining optical image resolution in the presence of a microparticle.

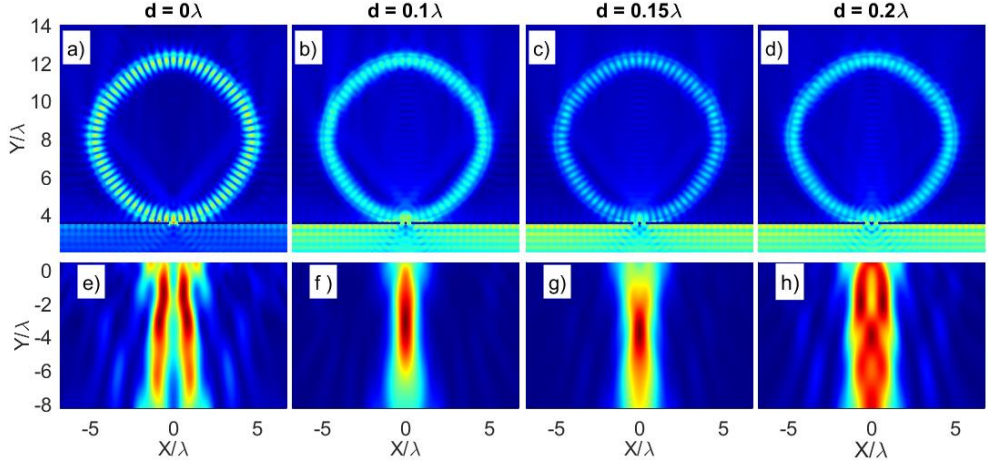

**Fig. 3.** (a, b, c, d) - Near-field distributions around the slits for R = 4.4λ, corresponding to the break points in Fig. 1(b), with slit distances d/λ = 0, 0.1, 0.15, 0.2, respectively. (e, f, g, h) - Image fields corresponding to the cases shown in the images above. Formal conditions for distinguishability are satisfied at d/λ = 0, however, in reality, the slits are distinguishable only at d/λ = 0.2. The maximum values of the near fields for (a)-(d) are related as 1:0.4:0.4:0.4.

The enhancement of the near field leads to an increased effect of secondary illumination of the sample by the field circulating inside the particle, which in turn results in the observed features. To verify this hypothesis, we performed resolution calculations where the source was modeled as two incoherent point dipoles represented as a uniformly distributed current within a cylinder with radius of 0.025λ. The calculation geometry and model exactly match those of the reflection mode case, meaning that we effectively replaced the reflecting objects with point sources. In this case, local field enhancement also occurs at the mentioned points; however, it does not affect the emitted field of the source. The optical resolution remains close to the theoretical limit of 0.5λ and does not exhibit anomalies at these points, as shown in Fig. 4(a).

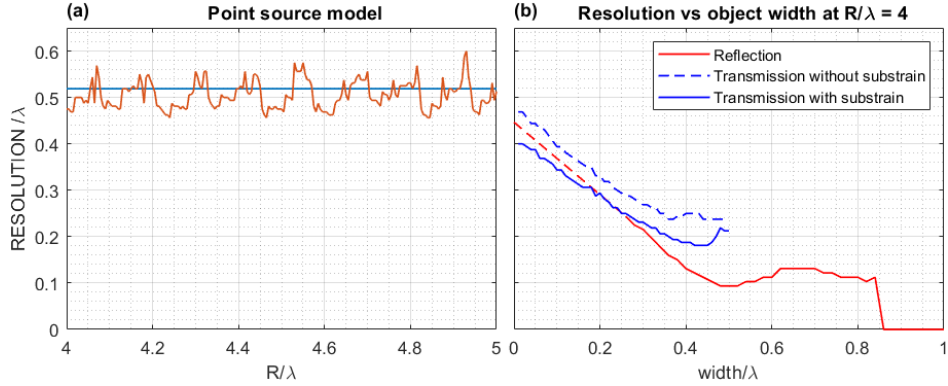

**Fig. 4.** (a) The dependence of optical resolution in the point source model for free space (blue line) and in the presence of a microparticle (red curve). The optical resolution in free space is 0.52λ, while in the case of a microparticle, it oscillates with a small amplitude around this value. (b) Dependence of optical resolution on the object width (plates or slits) for Köhler illumination. The red dashed line denotes the extrapolated behavior in the limit of vanishing panel width, as direct calculation is impeded by the poor contrast of the image. The solid and dashed blue curves represent analogous dependencies in the presence and absence of a substrate beneath the slits, respectively. (Will be updated upon further calculation)

It should be noted that to achieve a resolution of approximately 0.5λ in the point source model, the calculation domain was extended by 10λ along the x-axis. Otherwise, the resolution would be limited to 0.6λ. This difference is explained by the increased numerical aperture, which enhances the collection of field information. However, this extension does not affect the results for the microsphere, as most of the rays passing through the microsphere are confined within a limited cone [14]. Moreover, only the x-component of the field was used in image calculations.

A comparison of the results in Fig. 4(a) for the point source model and those in Fig. 1 for the simulation model indicates that the key factors enabling super-resolution are the interactions between radiation and the investigated sample, as it was noted to some extent in Ref. [10]. According to the presented 2D analysis, microparticle-assisted microscopy does not enhance optical resolution in terms of resolving individual *point sources* beyond the diffraction limit. A similar conclusion was drawn in our previous work [14], as well as in our theoretical analysis of optical resolution from the perspective of the limited number of modes excited within the microsphere [12]. This leads to the conclusion that the distinguishability of objects depends not only on the distance between them but also on their width, Fig. 4(b) illustrates this dependence at R/λ=4. The resolution tends to λ/2 when object width→0. For object sizes smaller λ/2, an approximate resolution criterion for a microparticle in the 2D case is given by:

$$d + w \approx \frac{\lambda}{2}, \qquad (1)$$

where $w$ is the characteristic size of the object under investigation, and $d$ is the minimal distance (resolution value) between adjacent objects. It should be noted that this estimate is qualitative. According to the data in Fig. 1, condition (1) is satisfied with good accuracy for the average value: $\langle d \rangle + w = 0.48\lambda$ in the reflection mode and $\langle d \rangle + w = 0.51\lambda$ in the transmission mode. We also note that this linear trend of resolution approaching λ/2 is valid only outside of resonant cases. For object sizes larger than λ/2, the optical resolution is almost independent of the object size. At R/λ = 4, it is approximately 0.12λ in the reflection mode and 0.21λ in the transmission mode. It should be noted that these values may vary for other R/λ ratios. The graph in Fig. 4(b) shows almost linear reduction of the resolution for reflection mode as the object width decreases. Unfortunately, performing calculations for reflection mode in the limit width→0 was not possible due to low contrast. However, interpolation of the results suggests a resolution of 0.44λ, which is only slightly different from the point source model 0.47λ for the same R/λ. In the transmission mode, when a highly conductive material is considered, this issue does not arise, and the calculations are performed down to slits width of 0.01λ. To ensure adequate spatial resolution for small slit widths, a mesh step of 0.001λ was used in the slit region. However, this limiting behavior is affected by the presence of a substrate beneath the slits. Under oblique illumination, the field distribution along the lower surface of the conductor exhibits a characteristic pattern of alternating maxima and minima. Antisymmetric modes are excited when the field has a maximum near one slit and a minimum near the other, resulting in a π phase difference between the transmitted waves. As a result, antisymmetric mode excitation occurs, producing an image with a characteristic two-lobed pattern [16] that enables the distinction between the studied structures. Since the distance between the field lobes depends on both the angle of incidence and the refractive index of the material, the ultimate resolution is determined by the refractive index of the substrate on which the slits are placed. Consequently, the resolution limit at width→0 in Fig. 4(b) for transmission mode is determined by the refractive index of the substrate.

## 3. Discussion

The optical resolution of a microparticle exhibits a nonlinear dependence on its size. This dependence may vary depending on the resolution criterion. In this study, we employ a method based on identifying the global maximum within a specified region around the geometric focus, which requires further clarification.

In the three-dimensional case, the resolvability of objects can be assessed by analyzing the intensity distribution in the focal plane. A structured intensity pattern that cannot result from random interference indicates the presence of the resolvable objects. Therefore, the focal plane should be selected to represent the geometry of the sample. In the two-dimensional case, with a fixed observation plane, only a one-dimensional field distribution $\mathbf{E}^{im}(x)$ can be recorded, making it challenging to directly correlate the observed intensity dip between maxima with the sample's geometry. If the geometry of the sample is known a priori, the focal line $y=y^{image}$ can be chosen to optimally represent its structure. However, in microscopy, where the goal is to investigate unknown objects with an undetermined geometry, such an approach is not applicable.

For example, depending on the choice of $y^{image}$, the sample may be interpreted either as a single slit or as two separate slits. The global maximum search eliminates this ambiguity. Furthermore, the definition of the optical resolution limit implies that if the distance between objects exceeds this threshold, they must remain distinguishable. If the focal plane is selected outside the global maximum, this condition may not be satisfied.

An alternative approach is to assess resolution at a fixed position of $y^{image}$, for example, in the region of the geometric focus. However, due to significant aberrations introduced by the microparticle, which cause all rays to converge at a single point, this choice becomes somewhat arbitrary. In this study, we aim to establish a resolution criterion that is both objective and robust.

Equation (1), which provides an estimate of the resolution limit, applies to the two-dimensional case; its validity in three dimensions has not been demonstrated so far. To confirm this result for the 3D case, one would need to perform calculations similar to those shown in Fig. 4(b), which is computationally challenging. The present work is limited to 2D simulations; however, the presented model can be directly extended to the 3D case. Thus, this study lays the groundwork for future research.

## 4. Materials and methods

### 4.1 Image field calculation approaches

There are several methods in the literature for calculating the image field at micrometer scales. In this section, we briefly describe them for the two-dimensional case. Let the field $\mathbf{E}$ be generated by a certain coherent monochromatic source, which we will refer to as the source field. The temporal dependence takes the form $\sim e^{-i\omega t}$, which we will omit unless stated otherwise. The source field can be associated with its corresponding image field $\mathbf{E}^{im}$ according to the following formula:

$$\mathbf{E}^{im}(r_0) = \frac{-i}{4} \int_\Gamma \left[ G(\mathbf{n}, \nabla)\mathbf{E}^* - \mathbf{E}^*(\mathbf{n}, \nabla)G \right] dl , \qquad (2)$$

where $\Gamma$ is an arbitrary curve homotopic to an infinite line, $G = H_0^{(1)}(k|r - r_0|)$ is the Hankel function of the first kind, and * denotes complex conjugation, $k=2\pi/\lambda$, $\nabla = \partial/\partial \mathbf{r}$, $\lambda$ is wavelength. The complex conjugation of the source field reverses the wave front, transforming the field from diverging to converging to the source. Equation (2) can be rewritten in Fourier space as follows:

$$\mathbf{E}^{im}(r_0) = \frac{1}{2\pi k} \int_{-k}^{k} \tilde{\mathbf{E}}^*(k_x, y_0) e^{-i(k_x x - k_y(y - y_0))} dk_x , \qquad (3)$$

where $\tilde{\mathbf{E}}$ is the Fourier transform of the field $\mathbf{E}$ at $y=y_0$, and $k_y = \sqrt{k^2 - k_x^2}$. According to Eq. (3) and in accordance with Abbe's definition, the image field does not contain evanescent harmonics with $k_y > k$.

In the case when the source field is generated by scattering on a microparticle, the fields $\mathbf{E}^{im}$ and $\mathbf{E}$ can be expanded in terms of cylindrical functions. For the TM geometry ($E_x=E_y=0$), these expansions take the following form [17]:

$$\mathbf{E} = \sum_{l=-\infty}^{\infty} a_l \mathbf{N}_l,$$
$$\mathbf{E}^{im} = \sum_{l=-\infty}^{\infty} a_l^{im} Rg\mathbf{N}_l^*. \quad (4)$$

For convenience, the field $\mathbf{E}^{im}$ is expanded using complex-conjugated functions, $\mathbf{N}_l = e^{il\varphi} H_l^{(1)}(k\rho)\mathbf{e}_z$, $H_l^{(1)}$ is the Hankel function of the first kind, and Rg denotes the extraction of the regular part of the expression. Due to the linearity of Eqs. (2)–(3), the column vectors of the expansion coefficients of the source field ($a_l$) and the image field ($a_l^{im}$) are related by a matrix equation of the form:

$$\begin{pmatrix} \cdots \\ a_{-1}^{im} \\ a_0^{im} \\ a_1^{im} \\ \cdots \end{pmatrix} = \mathbf{A}^{im} \begin{pmatrix} \cdots \\ a_{-1}^* \\ a_0^* \\ a_1^* \\ \cdots \end{pmatrix}. \quad (5)$$

The components of the matrix $\mathbf{A}^{im}$, which link the coefficient $a_n$ on the right-hand side with the coefficient $a_m^{im}$ on the left-hand side of the expression, are given by $(\mathbf{A}^{im})_{nm}=\text{sinc}((n-m)/2)$. This operator was first computed for the three-dimensional case in Ref. [12]. In the two-dimensional case, the calculations are analogous; for more details, see Supplementary.

For the TE geometry ($E_z = 0$), the fields $\mathbf{E}$ and $\mathbf{E}^{im}$ can be similarly expanded using functions $\mathbf{M} = \text{rot}(\mathbf{N})/k$. Due to the linearity of this transformation, Eq. (5) and the components of the matrix $\mathbf{A}^{im}$ remain unchanged.

If the source field is non-stationary, $\mathbf{E} = \mathbf{E}(t)$, there are two approaches to compute the corresponding time dependence of $\mathbf{E}^{im}(t)$. In the first approach, the time dependence can be expanded into a Fourier series. Then, by applying transformations (2)–(3) to each component and performing an inverse Fourier transform, the dependence $\mathbf{E}^{im}(t)$ can be obtained.

The second approach involves the FDTD (Finite-Difference Time-Domain) method [14]. In this method, the field sources are explicitly defined at the grid nodes, and the field propagation in space is computed using a finite-difference scheme in the time domain. To ensure that the generated pulse corresponds to the image field rather than the source field, the time dependence must be reversed.

For the TM geometry, the setup of sources for generating the image field is implemented as follows:

$$E_z^{im}(x, y = y_o, t+dt) = E_z^{im}(x, y = y_o, t) + $$
$$+ (E_z(x, y = y_o + dy, -t) - E_z(x, y = y_o, -t))/dy + E_z(x, y = y_o, -t)/dy, \quad (6)$$

$$E_z^{im}(x, y = y_o + dy, t+dt) = E_z^{im}(x, y = y_o + dy, t) - E_z(x, y = y_o, -t)/dy, \quad (7)$$

where $y=y_0$ is the line on which the sources are defined, and $dt$, $dy$ are the time and spatial grid steps, respectively. A similar approach can be applied in the TE geometry for each component $E_x$, $E_y$ or by defining the source through the magnetic component $H_z$.

In our calculations, we used the FEM method, which allows us to obtain the spatial distribution of the source field. For this reason, we applied Eq. (2). The curve $\Gamma$ was chosen as a smooth, bell-shaped contour enclosing both the particle and the surrounding space. Such configuration captures all rays emanating from the microsphere. If $\Gamma$ is chosen as a straight line, achieving the same result would require the line to be significantly larger than the microsphere's diameter, substantially increasing the simulation domain. Similar optimizations were performed in Ref. [9].

*4.2 Calculation Setup*

For accurate calculations, the mesh size in the source generation and object regions must be especially fine. The maximum distance between mesh nodes was set to 0.066λ, with further refinement in the specified regions to 0.005λ. Simulations were conducted for microsphere sizes R=(4:0.005:5)λ with a refractive index of n=1.46.

In the reflection geometry, the sample was composed of rectangular perfect conductors with a width of 0.25λ and a height of 0.1λ. For the transmission geometry, the sample was represented by slits in an opaque screen of the same dimensions. The entire structure was placed on a substrate with a refractive index of n=1.46.

The dimensions of the simulated region were (4.4R+8λ)×(2R+5.9λ). If the contact point between the microparticle and the substrate is taken as the origin, the curve Γ is defined by the following equation:

$$y = 3\lambda + (2R + 0.2\lambda)\left[1 + \left(\frac{x}{1.2R}\right)^4\right]^{-1}.$$

This dependence ensures a 3λ-offset from the substrate surface to avoid edge effects. The overview of the calculation setup is shown in Fig. 5.

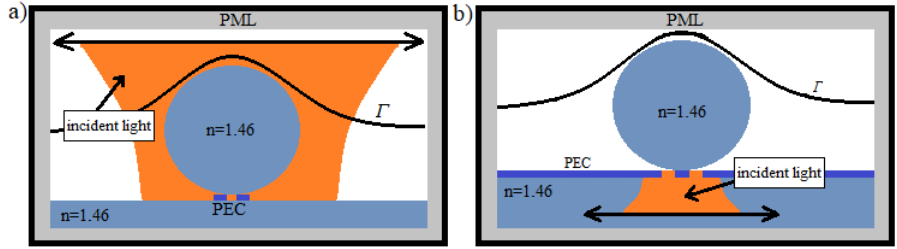

**Fig. 5.** General calculation scheme (not to scale): (a) reflection mode, (b) transmission mode.

To ensure a physically accurate assessment of the resolution, it is crucial to define properly the illumination conditions. In the calculations, we used the Köhler illumination scheme, where the sample is illuminated by incoherent plane waves at various incident angles. Plane-wave illumination is confined within a cone with a half-angle of 3π/8 (NA = 0.92), with an angular step of π/100, for both transmission and reflection geometries.

Fig. 6 shows the near-field and the corresponding image field in the absence of a microparticle at different object separation distances. The figure illustrates two representative scenarios: zero spacing between adjacent object elements and the minimal separation distance at which the object features remain just resolvable.

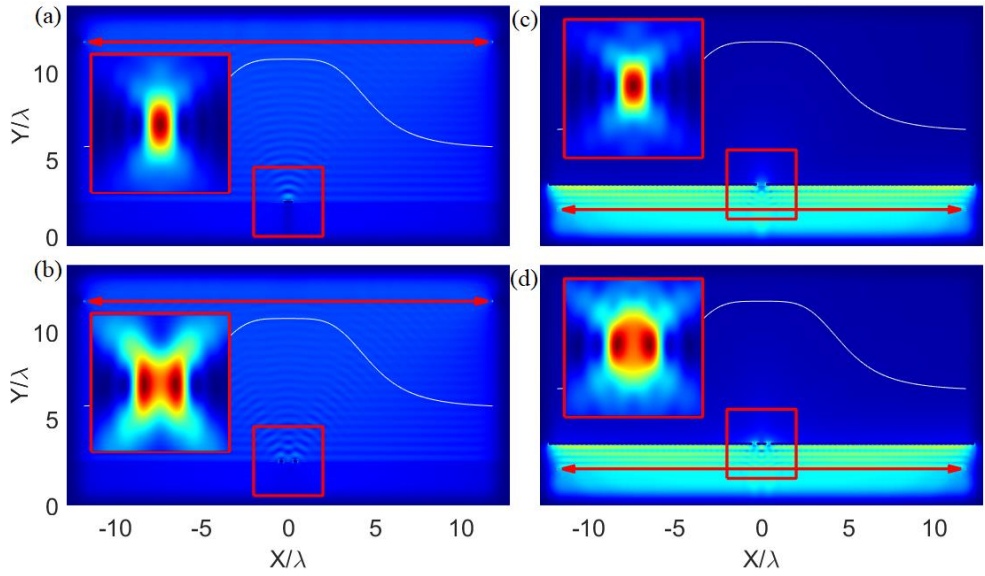

**Fig. 6.** Real and image fields without a microsphere. The white curve represents the integration path Γ, the red double arrow line represents source injection line. The insets display the image field calculated for the area highlighted by the red square. (a, b) – Reflection mode, with distances between objects d=0 and d=0.56λ, respectively. (c, d) – Transmission mode, with distances between apertures d=0 and d=0.57λ, respectively.

*4.3 Image formation algorithm by incoherent source and criterion of the resolution*

The algorithm for calculating the image field for a spatially incoherent source includes the following key steps: computing the image field for a specific plane wave incidence angle, varying the angle of incidence, and summing the intensity contributions from all angles. The image field is computed using Eq. (2) in a rectangular area centered at the geometric image position, determined by the magnification formula $n/(2-n)R$, with dimensions $(2.2R+4)\times 2R$.

The resolution condition is defined as follows: let the maximum of the image intensity field distribution for $x>0$ be at point $A_1$, and for $x<0$ at point $A_2$. The objects are considered resolvable if the intensity at the midpoint $A_{12}$ exhibits an 80% dip relative to the maxima at $A_1$ and $A_2$ points.

In many cases, due to overlapping image fields from each object, the maximum shifts toward the center, forming a single peak above or below two distinct local maxima. If this central maximum is excluded, the resolution criterion is satisfied, and the objects (e.g., slits) can be considered resolvable. However, such cases are not considered in the presented calculations, meaning that the optical resolution is estimated conservatively, or "from below".

Fig. 7 shows the near-field and the corresponding image field for a microparticle with R=4.5λ at different object distances for critical scheme illumination. The microparticle does not generate an image in the same manner as in free space. In the case of a microcylinder, the image appears as elongated bright stripes, whereas in free space, the image is localized on a wavelength scale, as evident from the comparison between Fig. 6 and Fig. 7.

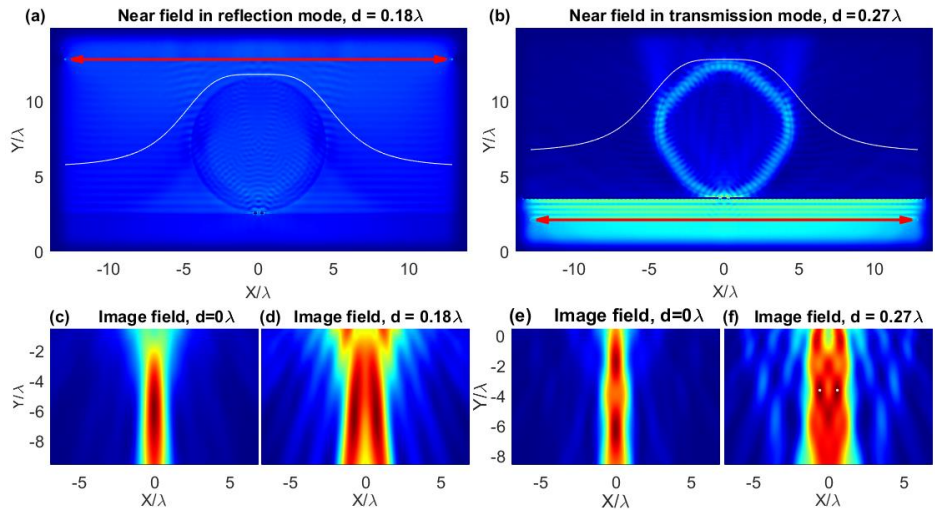

**Fig. 7.** (a, b) Near-field distribution in reflection and transmission modes, respectively. Image field in reflection mode with object separation of d=0λ (c), d=0.18λ (d). Image field in transmission mode with slit separation of d=0λ (e), d=0.27λ (f). The white dots in (f) indicate the maximum field. The particle radius is R=4.5λ, n = 1.46. The white curve represents the integration path Γ, the red double arrow line represents source injection line.

**Funding.** The study was carried out within the framework of the state assignment of the Lomonosov Moscow State University; the work was also supported by a grant from the Foundation for the Development of Theoretical Physics and Mathematics "BASIS" (grant No. 23-1-1-61). Z.W. is thankful for support from the Leverhulme Trust (RF-2022-659), the Royal Society (IEC\R2\202178), and Bangor University (BUIIA-S46910).
**Disclosures.** The authors declare no conflicts of interest.
**Data availability.** Data is available upon reasonable request.